\documentclass[aps,prl,twocolumn,showpacs,groupedaddress,superscriptaddress]{revtex4-1}

\usepackage{amsmath, amsthm, amssymb}
\usepackage{graphicx}
\usepackage{xcolor}
\definecolor{mydarkgreen}{rgb}{0.0,0.5,0.0}
\usepackage[linktocpage=true,
  colorlinks=true, 
  pdfborder={0 0 0},
  linkcolor=blue,
  citecolor=red,
  filecolor=yellow,
  urlcolor=mydarkgreen,
  bookmarks,
  pdftitle={Exact\ properties\ of\ density\ functionals\ for\ materials\ under\
pressure},
  pdfauthor={},
  pdfkeywords={DFT,scaling,virial,pressure,solids},
]{hyperref} 

\def\bea{\begin{eqnarray}}
\def\eea{\end{eqnarray}}
\def\ben{\begin{equation}}
\def\een{\end{equation}}
\def\benu{\begin{enumerate}}
\def\enu{\end{enumerate}}

\def\bei{\begin{itemize}}
\def\eei{\end{itemize}}
\def\beit{\begin{itemize}}
\def\eit{\end{itemize}}
\def\benu{\begin{enumerate}}
\def\enu{\end{enumerate}}

\def\veps{\varepsilon}

\def\sss{\scriptscriptstyle\rm}

\def\g{_\gamma}

\def\hatH{{\hat H}}
\def\1var{(\bx_1...\bx\N)}

\def\half{\frac{1}{2}}

\def\br{{\bf r}}

\def\bx{{x}}

\def\c{_{\sss C}}
\def\s{_{\sss S}}
\def\xc{_{\sss XC}}

\def\N{_{\sss N}}
\def\H{_{\sss H}}

\def\sph_int{ {\int d^3 r}}

\def\Eq#1{Eq.~(\ref{#1})}
\def\eq#1{(\ref{#1})}

\def\sig#1#2{}

\def\bay{\begin{array}}
\def\eay{\end{array}}
\def\beit{\begin{itemize}}
\def\eit{\end{itemize}}

\def\bx{{\bf x}}

\def\crap#1{The following is crap:\\ #1}
\def\crap#1{\bf !!!!Erroneous entry was removed here!!!!}
\def\bei{\begin{itemize}}
\def\eei{\end{itemize}}
\def\benu{\begin{enumerate}}
\def\enu{\end{enumerate}}

\def\s{_{\rm s}}

\def\N{\tilde{N}}
\def\0{^{(0)}}

\def\1{^{(1)}}
\def\0{^{(0)}}
\def\del{\partial}

\def\2F1{_2\mathrm{F}_1}

\newcommand{\brk}[3]{\langle #1\,|\ #2 \ |\,#3 \rangle}

\newcommand{\bra}[1]{\langle #1\,|}
\newcommand{\ket}[1]{|\,#1 \rangle}
\newcommand{\Bra}[1]{\left\langle #1\,\right|}
\newcommand{\Ket}[1]{\left|\,#1 \right\rangle}

\def\br{{\bf r}}

\def\co{(color online)  }
\def\br{{\bf r}}

\begin{document}

\title{The virial theorem and exact properties of density functionals for periodic systems}

\author{H. Mirhosseini}
\altaffiliation{Current address: Johannes Gutenberg University, 
55122 Mainz, Germany}
\affiliation{Max Planck Institute of Microstructure Physics, 
Weinberg 2, 06120 Halle (Saale), Germany}

\author{A. Cangi}
\affiliation{Max Planck Institute of Microstructure Physics, 
Weinberg 2, 06120 Halle (Saale), Germany}

\author{T. Baldsiefen}
\affiliation{Max Planck Institute of Microstructure Physics, 
Weinberg 2, 06120 Halle (Saale), Germany}

\author{A. Sanna}
\affiliation{Max Planck Institute of Microstructure Physics, 
Weinberg 2, 06120 Halle (Saale), Germany}

\author{C. R. Proetto}
\affiliation{Centro At\'omico Bariloche and Instituto
Balseiro, 8400 S.C. de Bariloche, R\'io Negro, Argentina}
\affiliation{Max Planck Institute of Microstructure Physics, 
Weinberg 2, 06120 Halle (Saale), Germany}

\author{E. K. U. Gross}
\affiliation{Max Planck Institute of Microstructure Physics, 
Weinberg 2, 06120 Halle (Saale), Germany}

\date{\today}

\begin{abstract}
In the framework of density functional theory, scaling and the virial theorem
are essential tools for deriving exact properties of density functionals.
Preexisting mathematical difficulties in deriving the virial theorem via scaling
for periodic systems are resolved via a particular scaling technique. 
This methodology is employed to derive universal properties
of the exchange-correlation energy functional for periodic systems.
\end{abstract}


\maketitle


Presently, Kohn-Sham (KS) density functional theory (DFT) \cite{HK64,KS65} is
the state-of-the-art \emph{ab initio} method for predicting the electronic
properties of materials due to its balance between accuracy and
computational efficiency. It relies on the mapping of the interacting many-body
system onto a noninteracting system of KS electrons that yields the
true density. This is achieved by introducing a local, one-body potential, the
KS potential, mimicking all interelectronic interactions via Hartree and
exchange-correlation (XC) contributions. Although being formally exact, in
practice the XC piece needs to be approximated. 
For electronic structure calculations of periodic systems, most
commonly, the local density approximation (LDA) \cite{KS65} or generalized gradient
approximations (GGAs)\cite{PBE96} are applied. 
Such calculations are performed either at zero or finite temperature \cite{M65,PPFS11}.

Nonempirical improvements upon these approximations rely on exact properties 
of the XC functional that provide guidance for constructing accurate approximations.  
But so far exact properties of the XC functional have only been derived 
for localized systems \cite{LP85}. 
As we demonstrate in this paper, some exact properties of the
XC functional change for periodic systems -- a fact that has been 
completely neglected for functional construction so far.
The quantum mechanical virial theorem (VT) 
and uniform coordinate scaling (UCS) have been essential mathematical
tools for deriving such exact properties for localized systems\cite{DG90}.

In quantum mechanics, the VT was derived in different ways \cite{MM07}. 
At zero temperature,  
within the Born-Oppenheimer approximation, for all Coulombic matter with the electronic Hamiltonian
\ben
\hat {H}^{\Omega_1} = \hat {T} + \hat{W} +\hat{V}^{\Omega_1}\,,~~~~~
\hat {H}^{\Omega_1}\, \Psi^{\Omega_1} = E^{\Omega_1}\, \Psi^{\Omega_1}\,, 
\een
and under the assumption of hydrostatic pressure,
the VT states that
\ben\label{VT}
2 T^{\Omega_1} + W^{\Omega_1} + V^{\Omega_1}
= -D~\Omega_1\, \left. \del_\Omega  E^\Omega\right|_{\Omega=\Omega_1}\,.
\een 
As it will be shown later, one cannot derive \Eq{VT} for periodic systems by
uniform coordinate scaling method \cite{LP85}. In this paper we derive \Eq{VT}, in
particular for periodic systems, by introducing and using uniform coordinate and
potential scaling (UCPS). 
In \Eq{VT}, $T^{\Omega_1}=\brk{\Psi^{\Omega_1}}{\hat T}{\Psi^{\Omega_1}}_{\Omega_1}$, 
$W^{\Omega_1}=\brk{\Psi^{\Omega_1}}{\hat W}{\Psi^{\Omega_1}}_{\Omega_1}$, 
and $V^{\Omega_1}=\brk{\Psi^{\Omega_1}}{\hat V^{\Omega_1}}{\Psi^{\Omega_1}}_{\Omega_1}$
denote the expectation values of the kinetic, interelectronic interaction, and
external potential energy operators. 
Antisymmetric wave functions $\Psi^{\Omega_1}$ are eigenstates 
of $\hat{H}^{\Omega_1}$ which is defined on volume $\Omega_1$.
The subscript $\Omega_1$ of expectation values indicates 
the volume in which the operators are evaluated, $D$ denotes 
the dimensionality of space \footnote{for brevity we use a short-hand notation 
for derivatives, such as $\del_\Omega=\del/\del\Omega$.}. 
This general form of the VT is valid for localized systems (atoms and molecules),
strictly confined systems (particles in a box with hard walls), 
and periodic systems (solids):
As an example consider diatomic molecules\cite{S33} for which 
the right-hand side (RHS) of
\Eq{VT} reduces to $\left.  -R_1\, \del_R E^R \right|_{R=R_1}$, 
where $R_1$ denotes the distance between the nuclei. 
For strictly confined systems\cite{CP51} the RHS of \Eq{VT} becomes 
$\left. -L_1\, \del_L E^L\right|_{L=L_1}$,
where $L_1$ denotes the distance between the walls.
For the homogeneous electron gas (HEG)\cite{A67}, 
a very crude approximation to a periodic system, 
the RHS of \Eq{VT} is 
$\left. -r_{s,1}\, \del_{r_s}E^{r_s}\right|_{r_s=r_{s,1}} $,
where $r_{s,1}$ is the radius of a sphere that contains one electron.
In the VT for a periodic system, 
which we address in this work, 
$\Omega_1$ is generally considered as the volume of the unit cell.   
In the case of localized systems the RHS of \Eq{VT}
is proportional to the force that keeps the nuclei away 
from their equilibrium positions, whereas for periodic systems 
the RHS of \Eq{VT} contains an additional contribution 
of kinetic and interelectronic interaction energy, 
a so-called surface term \cite{MM07}. 
In this paper we derive the most general form of the VT valid 
for periodic systems under the hydrostatic assumption. 
This is done via a scaling technique 
developed in the following that relies on UCS, 
which in turn was used to obtain the VT, 
but only for localized systems \cite{Z80, FC82}.

In UCS the $D$-dimensional position vectors of the electrons
are scaled as ${\br}_i \to \gamma\, {\br}_i$, whereas other length scales of the
system stay \emph{fixed}. This defines 
\ben
\Psi^{\Omega_1}\g({\br}_1,\dots,{\br}_N) =
\gamma^{D N/2}\,\Psi^{\Omega_1}(\gamma\,{\br}_1,\dots,\gamma\,{\br}_N),
\een
where the prefactor is determined by requiring the normalization of the scaled
wave function on the scaled volume $ \Omega\g=\gamma^{-D}\Omega_1$. 
Recall that for localized systems the normalization volume is taken as infinite ($\Omega_\infty$).
and is therefore not affected by scaling. 
Employing the extremum principle, 
\ben\label{RRVP}
\left. \del_\gamma  \brk{\Psi^{\Omega_\infty}\g}
{\hat H^{\Omega_\infty}}{\Psi^{\Omega_\infty}\g}_{\Omega_\infty}\right|_{\gamma=1} 
= 0\,,
\een
and considering the scaling of expectation values, 
$T^{\Omega_\infty}\g = \gamma^2 T^{\Omega_\infty}$, 
$W^{\Omega_\infty}\g = \gamma W^{\Omega_\infty}$, and 
$V^{\Omega_\infty}\g = \int d^Dr\,n^{\Omega_\infty}(\br)\, v^{\Omega_\infty}(\br/\gamma)$
yields the VT for localized systems, 
i.e., \Eq{VT} becomes
\ben\label{VT.loc}
2 T^{\Omega_\infty} + W^{\Omega_\infty} - \int_{\Omega_\infty} d^Dr\ 
n^{\Omega_\infty}(\br)\ \br\cdot\nabla v^{\Omega_\infty}(\br) = 0\ .
\een
But, as we will show, \Eq{RRVP} is not a valid starting point 
for deriving the VT for periodic systems. The problem of
deriving the general VT via UCS has also been pointed out elsewhere 
\cite{BTHS10, EFG12}. Despite that fact, just the VT for localized systems has been
used to derive exact properties of the XC
functional \cite{LP85}, upon which most nonempirical approximations rely. 

In this paper we 
(i) pinpoint the mathematical difficulties of deriving the VT
via UCS for periodic systems, 
(ii) consequently, introduce a scaling technique 
that resolves the mathematical issues of UCS and 
derive the most general form of the VT 
(iii) derive fundamental scaling relations 
that steer the construction of functional approximations,  
(iv) find that the adiabatic connection remains unchanged for periodic systems, and
(v) generalize the derived VT to finite temperature.

\begin{figure}[t!]
\centering
  \includegraphics[width=0.85\columnwidth]{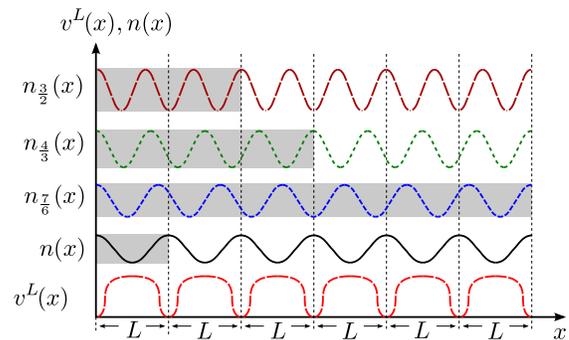} 
  \caption{{\co}~~ Sketch of coordinate-scaled densities on unscaled external
potential. Born-von Karman cells are denoted by the grey-shaded areas.}
  \label{fig.scaling}
\end{figure}

The key difference of localized versus periodic systems is 
in the treatment of the external potential. 
To show that we consider a scaling factor, 
arbitrarily close to 1, i.e., $\gamma_M=(M+1)/M$ with $M\in\mathbb{N}$ 
and $M \gg 1$.  For localized systems, $M$ can be chosen sufficiently large 
such that the difference between the scaled and unscaled
wave function becomes significant only at very large distances 
away from the center of mass of the atom or molecule 
not affecting the energy expectation value. Contrarily, this is generally
not valid anymore in the case of periodic systems 
where the expectation values are evaluated on a finite volume $\Omega_1$.
Scaling the wave function, then, defines a Born-von Karman cell of the size $ML$, 
where $L$ is the size of the chemical unit cell determined 
by the positions of the nuclei. This is shown for a one-dimensional system 
in Fig.~\ref{fig.scaling}.
The external potential energy per unit cell evaluated on scaled wave functions
then becomes
\begin{align}
  u^L_{\frac{M+1}{M}} &= \frac{M+1}{M^2}\int_0^{ML}dx\
n\left(\frac{M+1}{M}x\right)v^L(x)\ .
\label{extpuc1}
\end{align}
Considering a particular unit cell (denoted by index $i$), 
the electronic density with scaled argument $n(x(M+1)/M)$ 
is related to a density with an appropriately shifted argument $n(x+iL/M)$; 
by construction, these densities coincide at one border of the unit cell 
and their overall difference is of the order of $1/M$. 
Therefore the external potential energy per unit cell is 
\begin{align}
u^L_{\frac{M+1}{M}} 
&= \frac{M+1}{ML}\sum_{i=1}^M\frac{L}{M}\int_0^{L}dx\
n\left(x+i\frac{L}{M}\right)v^L(x)
\label{extpuc2}
\end{align}
up to corrections of order $\mathcal{O}(1/M)$.
In the limit $M\rightarrow\infty$ the sum becomes an integral 
and 
\begin{align}
  \lim_{M\rightarrow\infty} u^L_{\frac{M+1}{M}} 
&= \overline{n}\int_0^L dx\ v^L(x),\label{eq.u.lim}
\end{align}
where $\overline{n}$ is the average density. In general Eq. \eq{eq.u.lim} is not
equal to the expectation value of the external potential  evaluated on the
unscaled wave function, i.e., while the kinetic and interelectronic interaction energy 
change smoothly with $\gamma$, the external potential energy and consequently 
the total energy are discontinuous at $\gamma = 1$. 
This poses a problem, because it implies that
\begin{align}
\left. \del_\gamma   E^{\Omega_1}_{\gamma, \scriptscriptstyle\rm UCS} \right|_{\gamma=1}
= \left. \del_\gamma 
\Bra{\Psi^{\Omega_1}_\gamma}\hat H^{\Omega_1}\Ket{\Psi^{\Omega_1}_\gamma}_{\Omega_1}
\right|_{\gamma=1} /M
\end{align}
is an illegitimate starting point for deriving the VT in the case of periodic systems.
This problem shows up every time an $\br$ operator appears as in \Eq{VT.loc}, 
making integration ill-defined for periodic systems -- 
a well-known fact that has also been addressed 
in the modern theory of polarization\cite{R94}.

To cure this problem, we introduce the methodology 
of uniform coordinate and potential scaling (UCPS) 
under which we recover the differentiability of 
$E^{\Omega_1}_{\gamma, \scriptscriptstyle\rm UCS}$ at
$\gamma=1$ essentially by scaling the external potential $\hat{V}^{\Omega_1}$. 
In detail, UCPS means the following: 
the electronic coordinate and wave function change according to UCS. 
Accordingly the external potential is scaled such
that its periodicity coincides with the scaled wave function, 
$\hat{V}^{\Omega_1}\rightarrow \hat{V}^{\Omega\g}$. 
The periodicity of a scaled wave function and the scaled external potential 
coincide and consequently \Eq{extpuc1} is a smooth function of $\gamma$.
\footnote{Inserting the correctly scaled external potential 
$v^{L\g}(x)$ into \Eq{extpuc1} and taking the limit $M\to\infty$, 
it can be proven that the discontinuity
in the external potential energy per unit cell disappears.}
It is useful to translate the concept of scaling to operators.
The identity
\begin{align}
\Bra{\Psi^{\Omega_1}\g}\hat O\g\Ket{\Psi^{\Omega_1}\g}_{\Omega\g}
&=\Bra{\Psi^{\Omega_1}_1}\hat O\Ket{\Psi^{\Omega_1}_1}_{\Omega_1} \label{eq.sca.id}
\end{align}
defines a scaled operator $\hat O\g$, where we
denote unscaled ($\gamma=1$) 
quantities explicitly by a subscript.
The scaled operators for the kinetic 
and interelectronic interaction energy
are simply related to their unscaled counterparts via
\begin{align}
	\hat T_\gamma &= \hat T/\gamma^2,
&	\hat W_\gamma &= \hat W/\gamma \label{eq.scale.tw}\ .
\end{align}
The spatial kernel of the external potential operator 
scales according to 
$v^{\Omega\g}\g({\br})=v^{\Omega\g}(\gamma\, \br)$.

We now apply UCPS and obtain a well-defined expectation value
\begin{align}
  E^{\Omega_1}\g&
=\Bra{\Psi^{\Omega_1}\g}\hatH^{\Omega\g}\Ket{\Psi^{\Omega_1}\g}_{\Omega\g}
=\Bra{\Psi^{\Omega_1}_1}\hat H_{1/\gamma}^{\Omega\g}\Ket{\Psi^{\Omega_1}_1}_{\Omega_1}
\label{eq.ev_gamma_2},
\end{align}
where the last equality follows from Eq. \eq{eq.sca.id}.
Due to the scaling of the external potential the derivative with respect to
$\gamma$ \emph{does} now exist at $\gamma=1$, but, in contrast to the case of localized
systems, it does \emph{not} vanish in general. This is due to the fact that
$\Psi_\gamma^{\Omega_1}$ is defined on a different volume $\Omega\g$ for each $\gamma$ and
therefore the extremum principle cannot be applied. However, we can relate
the derivative with respect to the scale parameter to the pressure $P$ of the
system:
\ben\label{eq.p}
 -P = \lim_{\veps\rightarrow 0}\frac{1}{\veps}
  \left(E_1^{\Omega_1+\veps} - E_1^{\Omega_1} \right)\,,
\een
where 
$E_1^{\Omega_1+\veps} = \Bra{\Psi_1^{\Omega_1+\veps}}
\hat H^{\Omega_1+\veps}\Ket{\Psi_1^{\Omega_1+\veps}}_{\Omega_1+\veps}$
and
$E_1^{\Omega_1} = \Bra{\Psi_1^{\Omega_1}}\hat H^{\Omega_1}
\Ket{\Psi_1^{\Omega_1}}_{\Omega_1}$. 
Since $\Psi_1^{\Omega_1+\veps}$ and $\Psi_1^{\Omega_1}$ 
are defined on different volumes, this complicates the use of perturbation theory. 
A way out of this dilemma is found by applying Eq. \eq{eq.sca.id} 
to $E_1^{\Omega_1+\veps}$ with the scale factor
\begin{align}
  \widetilde\gamma&=\left[(\Omega_1+\veps)/(\Omega_1)\right]^D.
\end{align}
Then, $E_1^{\Omega_1+\veps}$ can be calculated as the first order correction to
$E_1^{\Omega_1}$ under the perturbation 
$\Delta\hat H=\hat H^{\Omega_{\widetilde\gamma}}_{\widetilde\gamma}-\hat H^{\Omega_1}_1$.
Since we have ensured that the first order derivative with respect to $\gamma$
exists, we find 
\begin{align}\label{eq.ucvps.pres}
 \left. \del_\gamma E^{\Omega_1}_\gamma\right|_{\gamma=1}
&= -D\, \Omega_1\left. \del_\Omega  E^\Omega_1\right|_{\Omega=\Omega_1}\ .
\end{align}
Alternatively, this can be written as
\begin{align}\label{eq.vt.surf}
\begin{split}
& 2T^{\Omega_1} + W^{\Omega_1} 
+\int_{\Omega_1}d^Dr \ n^{\Omega_1}({\br})\, \del_\gamma v^{\Omega\g}({\br}/\gamma)|_{\gamma=1}\\
& = -D\, \Omega_1\, \del_\Omega E_1^\Omega |_{\Omega=\Omega_1}\,,
\end{split}
\end{align}
which reduces to \Eq{VT} for Coulombic matter.
Both Eqs.~(\ref{eq.ucvps.pres}) and (\ref{eq.vt.surf}), 
relating the change of the energy under a change of volume 
with a change in the scale parameter, 
yield the most general expression for the VT.  
This is one of our main results.   


We demonstrate the consistency of the VT for periodic systems that we
just derived with an elementary example of a solid explicitly.
Consider the simplified Kronig-Penney model \cite{PEH91}  -- a one-dimensional
lattice of Dirac delta functions of strength $\alpha$ separated by a distance
$L$ -- given by the Hamiltonian 
\begin{align}
H(x) &= -\half \del^2_x   - \frac{\alpha}{L}\sum_\nu
\delta\left(x-L\left(\nu-1/2\right)\right)\ .
\end{align}
A simple solution for positive energies is $\phi(x)\propto\cos(qx/L)$, 
where $q$ is determined from $q=q\cos(q)-\alpha\sin(q)$.
For a single particle in this state the energy is
\begin{align}
E^L_1 &= q^2/(2L^2)\ .\label{eq.kp.e}
\end{align}
The expectation values of the scaled kinetic and potential energy 
are related to the unscaled quantities simply by
\begin{align}
\Bra{\phi^{L}_{\gamma}}\hat T\Ket{\phi^{L}_{\gamma}}_{L\g}
&=\gamma^2 \Bra{\phi^{L}_{1}}\hat T\Ket{\phi^{L}_{1}}_L\label{eq.kp.t.sca}\,,\\
\Bra{\phi^{L}_{\gamma}}\hat V^{L\g}\Ket{\phi_\gamma}_{L\g}
&=\gamma^2 \Bra{\phi^{L}_{1}}\hat V^{L}\Ket{\phi^{L}_{1}}_L
\label{eq.kp.v.sca}\ .
\end{align}
Due to the specific form of the external potential there is a quadratic
dependence on the scaling parameter relating the scaled and unscaled potential
energy. Now we explicitly check Eq.~\eq{eq.ucvps.pres}. With Eqs.~\eq{eq.kp.e},
\eq{eq.kp.t.sca}, and \eq{eq.kp.v.sca}, the left-hand side yields
\ben
  \left. \del_\gamma E^{L}_\gamma\right|_{\gamma=1}
= q^2/L^2\label{kp.e.sca}\ .
\een
Using Eq. \eq{eq.kp.e}, the RHS of \Eq{eq.ucvps.pres} is then
simply shown to be identical to \Eq{kp.e.sca}.

In the framework of DFT, as was mentioned before, only the VT for localized
systems has been considered. Equipped with the new technique we are able 
to derive the exact properties of the XC functional valid for periodic systems. 
We apply \Eq{eq.ucvps.pres} to an interacting and a noninteracting system (KS
system) of the same density. Taking the difference of two VTs and thereby expressing
the interelectronic interaction in terms of KS quantities, i.e., 
$W^{\Omega_1} = U^{\Omega_1} + E\xc^{\Omega_1} - T\c^{\Omega_1}$
yields:
\begin{align}\label{hk-ks}
\begin{split}
& T\c^{\Omega_1} + U^{\Omega_1} + E\xc^{\Omega_1}  
 + D\, \Omega_1\, \del_\Omega\left.\left(E_1^\Omega-E\s^\Omega \right)\right|_{\Omega=\Omega_1}=\\
& - \int_{\Omega_1}\!\!\! d^Dr\ n^{\Omega_1}({\br}) 
 \del_\gamma \left.\left[ v^{\Omega_\gamma} \left(\frac{{\br}}{\gamma}\right) 
 - v\s^{\Omega_\gamma}\left(\frac{{\br}}{\gamma}\right) \right]\right|_{\gamma=1}\,,
\end{split}
\end{align}
where $U^{\Omega_1}$ denotes the Hartree, 
$E\xc^{\Omega_1}$ the XC, and $T\c^{\Omega_1}=T^{\Omega_1}-T\s^{\Omega_1}$ 
the kinetic correlation energies. 
The KS and external potential are scaled along the lines of \Eq{eq.sca.id} and
\ben\label{ks.pot.scaled}
v\s^{\Omega_1} ({\br}) - v^{\Omega_1}({\br}) = v\xc^{\Omega_1}({\br}) + v\H^{\Omega_1}({\br}),
\een
where $v\xc^{\Omega_1}({\br}) = \delta E\xc^{\Omega_1}/\delta n({\br})$ denotes the XC potential 
and 
$v\H({\br}) = \int_{\Omega_\infty}  d^Dr'\ n^{\Omega_1}({\br}')/|{\br}-{\br}'|$
the Hartree potential.
With \Eq{ks.pot.scaled} and using the fact that all terms containing Hartree and exchange
contributions cancel each other, we obtain the following virial relation for
the kinetic correlation energy:

\begin{align}\label{vt.tc}
\begin{split}
T\c^{\Omega_1} &= -E\c^{\Omega_1} 
+ \int_{\Omega_1}d^Dr \ n^{\Omega_1}({\br})\, \del_\gamma v\c^{\Omega\g}({\br}/\gamma)|_{\gamma=1}\\
&- D \Omega_1 \del_\Omega\left[ E\c^{\Omega}
-\int_{\Omega}d^Dr \ n^{\Omega}({\br})\, v\c^{\Omega}({\br})  \right]|_{\Omega=\Omega_1}\ .
\end{split}
\end{align}
The analysis of the slowly varying limit of Eq.~\eq{vt.tc} sheds some light on the differences of the
present work with the previous ones. For this, we need to use that 
$\del_\gamma v\c^{\Omega\g}({\br}/\gamma)|_{\gamma=1} 
\approx \del_\gamma v\c^{\Omega\g}({\br})|_{\gamma=1} 
= - D \, \Omega_1 \del_{\Omega} v\c^{\Omega}({\br})|_{\Omega=\Omega_1}$,
which is exact for the HEG, and approximately valid for systems with a slowly varying density. 
In this limit, Eq.~\eq{vt.tc}
may be accordingly expressed as
\begin{align}\label{svl}
\begin{split}
T\c^{\Omega_1} \approx 
& - E\c^{\Omega_1} - D \Omega_1 \del_\Omega E\c^{\Omega}|_{\Omega=\Omega_1} \\ 
& + D \Omega_1 \left.\left[ \del_\Omega \int_{\Omega} d^Dr \ n^{\Omega_1}({\br}) \, 
v\c^{\Omega_1}({\br}) \right] \right|_{\Omega=\Omega_1} \\
& + D \Omega_1 \int_{\Omega_1} d^Dr \ \left[ \del_{\Omega} 
n^{\Omega}({\br})|_{\Omega=\Omega_1} \right] \, v\c^{\Omega_1}({\br})  \ . 
\end{split}
\end{align}
For the HEG case, $n^{\Omega}({\br}) = n^{\Omega} = N/\Omega$, and
$v\c^{\Omega}({\br}) = v\c^{\Omega} = v\c(n^{\Omega})$; the last two terms on the RHS cancel with each other,
while the second term may be expressed as in Eq.~\eq{VT}, using that $\Omega_1 = 4 \pi r_s^3 / 3N$.  
For the evaluation of Eq.~\eq{svl} in the LDA, one needs to consider that 
$E\c^{\Omega} = \int_{\Omega} d^Dr\, n^{\Omega}({\br})\, \varepsilon\c[n^{\Omega}({\br})]$,
and that $v\c^{\Omega}({\br}) = v\c[n^{\Omega}({\br})]$. Proceeding along the lines of
Ref.~\cite{LP01}, we obtain the following well-known expression of Levy and Perdew (LP)\cite{LP85},
\ben\label{LP}
T\c^{\Omega_1} \approx - 4 \, E\c^{\Omega_1} + 3 \int_{\Omega_1} d^3r\ n^{\Omega_1}({\br})\, 
v\c[n^{\Omega_1}({\br})] \; .  
\een
Eq.~\eq{LP}, whose local version reads
$t\c[n^{\Omega_1}({\br})] = - 4\, \varepsilon\c[n^{\Omega_1}({\br})] + 3\, v\c[n^{\Omega_1}({\br})]$,
has been obtained in Ref.~\cite{LP85} restricting
the analysis to the case of localized systems, where, as discussed above, 
the normalization volume can be taken as
$\Omega_{\infty}$ and then is not affected by scaling. Here, proceeding from the \emph{extended} 
or \emph{periodic} scenario, we have arrived to the same result. This is however reasonable, since the
distinction between a system as \emph{localized} or \emph{extended} becomes progressively less clear as
the system approaches the truly slowly varying limit. Note however, that the HEG limit cannot be reached
under the assumptions of Ref.~\cite{LP85}, while it is exactly reproduced by our general approach.

\begin{table}[htb]
\caption{Numerical values for the kinetic correlation energy $T\c$ in \Eq{vt.tc}, 
computed for a set of realistic periodic systems\cite{compdetails} in LDA and GGA.
All values are given in Rydbergs/formula unit. 
$\Delta T\c$ is the difference between this exact form 
and the approximate one derived by Levy and Perdew (\Eq{LP}) 
evaluated on LDA quantities (energies, densities, and potentials). 
The already excellent agreement further improves (see $\Delta T\c^*$) 
by including GGA corrections on $v\c$ using the PBE XC functional 
(this difference is of the same order of magnitude of the estimated numerical accuracy 
of the calculations and therefore should be read as zero).}
\label{t:130312.EX.exp1.0.sft1.0}
\begin{ruledtabular}
\begin{tabular}{c c c c c c c}
& \multicolumn{2}{c}{$T\c$} 
& \multicolumn{2}{c}{$\Delta T\c/10^{-2}$}   
& \multicolumn{2}{c}{$\Delta T\c^*/10^{-5}$} \\
\cline{2-3}\cline{4-5}\cline{6-7}
pressure &   --  & 200GPa& --    & 200GPa&      -- &200GPa\\
\hline
Diamond  & 12.65 & 13.83 & -2.12 & -1.49 &    8.46 & 11.2 \\  
LiF      & 10.99 & 14.14 & -1.19 & -1.53 &    8.20 & 10.5 \\  
Graphite &  3.80 &  4.62 & -0.99 & -1.10 &   -0.31 &-0.46 \\  
LiFeAs   &  4.65 &  4.78 & -0.20 & -0.29 &    3.36 & 3.41 \\  
Ar       &  4.62 &  5.56 & -0.58 & -0.83 &    4.17 & 5.01 \\  
PdH      &  9.78 & 10.49 & -1.33 & -1.83 &   26.3  & 27.6 \\  
NaCl     & 14.57 & 20.29 & -1.13 & -2.36 &    5.17 & 7.29     
\end{tabular}
\label{tab:numerics}
\end{ruledtabular}
\end{table}

The expression in \Eq{vt.tc} for the kinetic correlation energy derived in this work is formally exact 
and equally valid for extended and localized systems, for both slowly and rapidly 
varying densities. 
We compare the exact expression in \Eq{vt.tc} with the LP simplified form 
given in \Eq{LP} by computing their difference 
for a set of real crystals of different chemical properties 
at low and high pressure\cite{compdetails}. 
In Tab.~\ref{tab:numerics} we evaluate the difference between 
Eqs.~(\ref{vt.tc}) and (\ref{LP}) on LDA ($\Delta T\c$) and 
GGA ($\Delta T\c^*$) quantities (energies, densities, and potentials).
As shown in Tab.~\ref{tab:numerics}, the difference within LDA is very small, 
of the order of $10^{-2}$ Ry per formula unit. 
This difference is hardly relevant for chemical application, 
and does not increase even when high pressure is applied.
When we turn to the GGA results, the difference in $T\c$ 
goes further down, by two orders of magnitude  
(below the estimated numerical error). This means that just by including 
the gradient corrections to $v\c$ the LP formula gives essentially the exact $T\c$.
Note however, that according to Eq.~(9) in Ref.~\cite{LP01}, the correct GGA for 
the kinetic correlation energy has more contributions than just those obtained from 
replacing $E\c$ and $v\c$ by the corresponding GGA quantities in \Eq{LP}.

We note in passing that the very important adiabatic connection formula\cite{PY89}, 
which gives the XC energy functional as a coupling-constant integral 
of the coupling-constant dependent expectation value of the interelectronic interaction 
($W$ in Eq.(1)), remains unchanged for periodic systems, 
since the adiabatic coupling-constant technique employed in its derivation 
does not change the periodicity of the density and Hamiltonian. 
This is consistent with the fact that the coupling-constant wave function 
may be expressed as $\Psi^{\Omega_1}\g[n_{1/\gamma}]$, 
which does not leave the domain of the Hamiltonian.

Eq. \eq{eq.ucvps.pres} is valid not only for the ground state, but for all
eigenstates $\Psi^{\Omega_1}_i$ of $\hat H^{\Omega_1}$. 
This enables us to derive corresponding versions of Eq. \eq{eq.ucvps.pres} 
for canonical and grand-canonical ensembles in the following.

Considering the canonical ensemble first, the equilibrium is defined as the
state with minimal free energy $F^{\Omega_1}=E^{\Omega_1}-1/\beta S^{\Omega_1}$, 
where $S^{\Omega_1}$ is the entropy and $\beta=1/(k_B\tau)$ is a measure 
for the temperature $\tau$, $k_B$ being Boltzmann's constant. A general quantum state 
is described by a statistical density operator $\hat \Gamma^{\Omega_1}$, 
a weighted sum of projection operators on the underlying Hilbert space
$\hat \Gamma^{\Omega_1}=\sum_i w^{\Omega_1}_i\ket{\Psi^{\Omega_1}_i}
\bra{\Psi^{\Omega_1}_i},\ (w^{\Omega_1}_i>0,\sum_i w^{\Omega_1}_i=1)$. 
The minimizing weights are then given by $w^{\Omega_1}_i=e^{-\beta E^{\Omega_1}_i}/Z$, 
where $E^{\Omega_1}_i$ is the i-th eigenvalue of $\hat H^{\Omega_1}$ 
and $Z$ is the normalization constant, i.e., the partition function.
This, in connection with Eq. \eq{eq.ev_gamma_2}, 
leads to the following definition for the free energy in UCPS:
\begin{align}
  F_\gamma^{\Omega_1}&=\sum_i\left(w^{\Omega_1}_i\bra{\Psi^{\Omega_1}_{i\ \gamma}}
\hat H^{\Omega\g}\ket{\Psi^{\Omega_1}_{i\ \gamma}}+\ln(w^{\Omega_1}_i)\right)\label{eq.free}
\end{align}
A coordinate scaling of the wave functions does not affect the weights
$w^{\Omega_1}_i$ and therefore leaves the entropic contribution invariant.
Furthermore, Eq. \eq{eq.free}, by definition, is minimal for the particular
choice of weights. The derivative with respect to volume therefore only
yields contributions from the volume dependence of the energy expectation value.
Combining these two findings we are lead to
\begin{align}\label{eq.ucvps.free}
\left. \del_\gamma F^{\Omega_1}_\gamma\right|_{\gamma=1}
&= -D\, \Omega_1\left. \del_\Omega  F^\Omega_1\right|_{\Omega=\Omega_1},
\end{align}
which is the equivalent of Eq. \eq{eq.ucvps.pres} for canonical ensembles. 

The same arguments can also be applied to the case of grand canonical ensembles and
its main thermodynamic variable, the grand potential 
$\Phi^{\Omega_1}=E^{\Omega_1}-\mu N-1/\beta S^{\Omega_1}$,
where the additional coupling to a particle bath is governed by the chemical
potential $\mu$, $N$ being the particle number,
\begin{align}\label{eq.ucvps.grand}
\left. \del_\gamma \Phi^{\Omega_1}_\gamma\right|_{\gamma=1}
&= -D\, \Omega_1\left. \del_\Omega  \Phi^\Omega_1\right|_{\Omega=\Omega_1}.
\end{align}

In this work we present the theoretical formalism of uniform coordinate 
and potential scaling in order to tackle a long-standing problem in DFT:
the formulation of a correct VT valid both for molecular (localized) systems
and for infinite periodic solids.
However, our numerical implementation and calculation for a set of realisitic 
periodic systems shows that corrections by our exact formulation are extremely small. 
And, hence, the localized form of the VT in the slowly-varying limit is sufficiently 
accurate for solid state applications.
Still there could be exotic cases in which the corrections become relevant.
Moreover, our scaling technique may find application in describing properties
of extended periodic systems at finite temperature, such as phase transitions.

We acknowledge useful discussions with S. Pittalis.
C.R.P. thanks \emph{CONICET} for partial financial support and 
\emph{ANPCyT} under grant number PICT-2012-0379.


\bibliographystyle{apsrev4-1}
\bibliography{refs}

\begin{thebibliography}{23}
\bibitem [{com()}]{compdetails}%
  \BibitemOpen
  \href@noop {} {}\bibinfo {note} {All systems are computed at the LDA relaxed
  structure. The numerical error in $T\c$ due to the numerical derivatives with
  respect to $\gamma$ has been estimated to be of the order of $10^{-4}$ Ry.
  Calculations have been done in the norm conserving pseudopotential
  approximation within the ESPRESSO planewave code\cite{QE09}. The planewave
  expansion of the Block orbitals has been cut at 100 Ry or above.}\BibitemShut
  {Stop}%
\end{thebibliography}

\end{document}